# Cross-linking of polyesters based on fatty acids

Sylwia Dworakowska[1,2], Cédric Le Coz[2], Guillaume Chollet[3], Etienne Grau*[,2], Henri Cramail*[,2]

[1] Cracow University of Technology, Faculty of Chemical Engineering and Technology, Warszawska 24, 31-155 Cracow, Poland

[2] Univ. Bordeaux, CNRS, Bordeaux INP, LCPO, UMR 5629, F-33600, Pessac, France

[3] ITERG, Lipochimie Hall Industriel, 11 rue Gaspard Monge, 33600 Pessac Cedex, France

*Correspondence: (Etienne Grau, Henri Cramail, Laboratoire de Chimie des Polymères Organiques, Université de Bordeaux, UMR5629, CNRS - Bordeaux INP - ENSCBP, 16 Avenue Pey-Berland, 33607 Pessac Cedex, France ; **Fax:** +33 05 40 00 84 87 **Email:** egrau@enscbp.fr, cramail@enscbp.fr)

## Abstract

This paper aimed at obtaining cross-linked polymeric materials of biomass origin. For this purpose one-pot polyesterification of methyl ricinoleate and methyl 12-hydroxystearate using titanium isopropoxide as a catalyst has been performed leading to polyesters known as estolides. The obtained estolides were successfully cross-linked using dicumyl peroxide or a sulfur vulcanization system. The so-formed bio-based elastomers appeared to exhibit promising properties. The latter were analyzed by mechanical tensile tests and thermal techniques (TGA, DSC, DMA) and showed high thermal stability ($T_{5\%}$ = 205–318 °C) and tailored physico-mechanical properties (low glass transition temperature in the range from - 69 to -54 °C) and good tensile strength (0.11–0.40 MPa). Networks prepared from high molecular weight estolides appear to be promising bio-based elastomers.

**Practical applications**

The vegetable oil-based estolides described in this contribution, are new fully bio-based precursors for further elastomers synthesis. The resulting estolide networks (obtained by peroxide or sulfur cross-linking) exhibit tailored thermo-mechanical properties.

**Graphical Abstract**

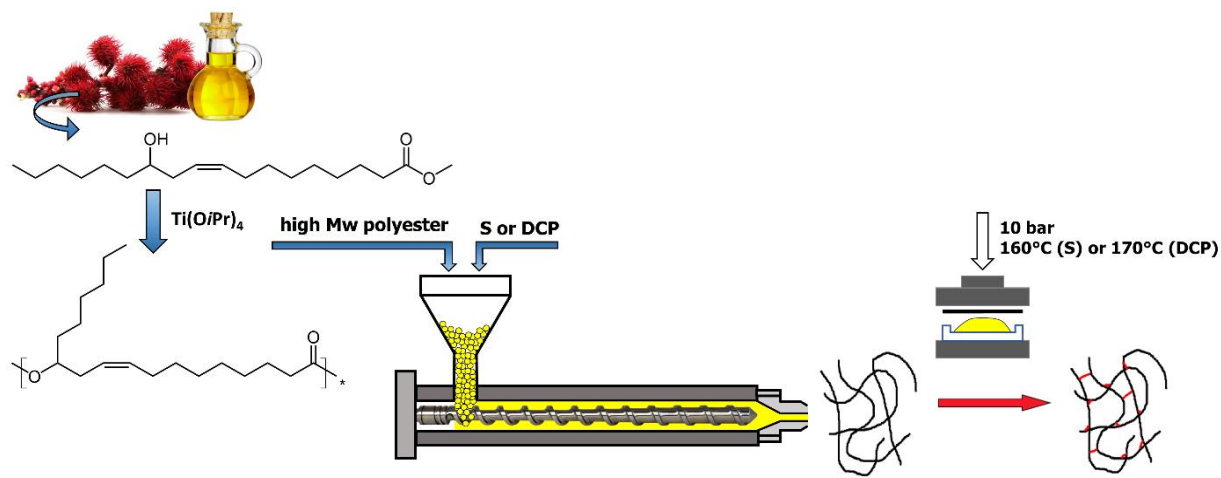

## 1. Introduction

Since the discovery of the vulcanization process of natural rubber in 1839 by Charles Goodyear [1] it is observed a continuous growth of the elastomer technology [2]. Nowadays, elastomers represent a vast market and are widely used in the mass production of components such as tires and non-tire applications for automotive, building, wires, footwear and many others [3]. According to the International Rubber Study Group (IRSG), the world elastomer production reached 28 MT in 2017, whereas synthetic elastomer occupied more than 15 MT [4].

Elastomers are usually prepared through the cross-linking of low Tg pre-polymers (precursors) through sulfur vulcanization or peroxide curing [5]. Among peroxides, dicumyl peroxide appears to be one of the most widely used cross-linking agents [6]. Indeed, this radical cross-linking agent enables the formation of the very stable C–C single bonds as shown in Scheme 1 [7]. In this case, two mechanisms are described: hydrogen abstraction (for saturated precursors) and/or radical addition (for unsaturated ones) (Scheme 1).

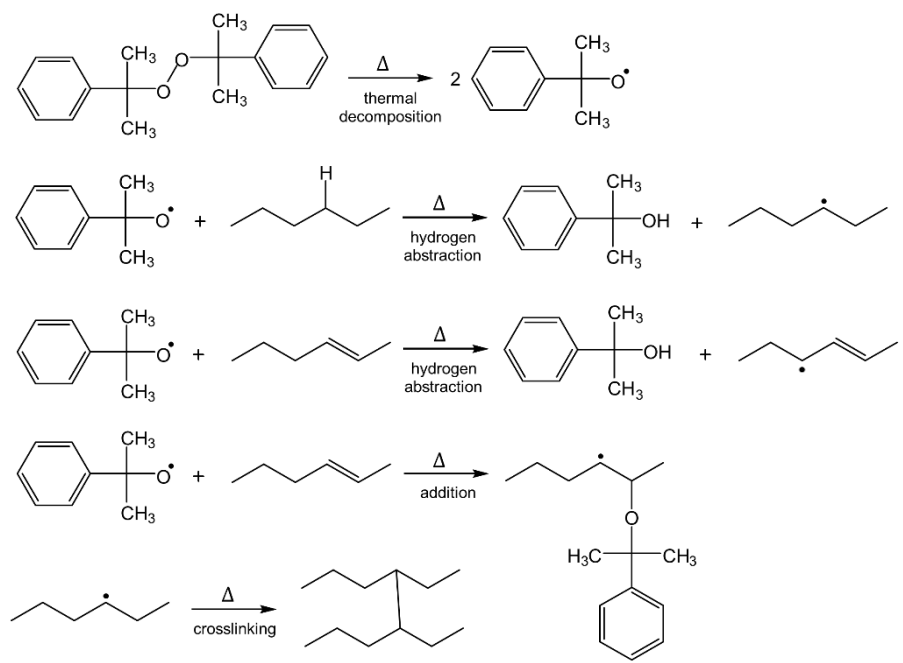

**Scheme 1.** Mechanism of radical cross-linking of saturated and unsaturated precursors using dicumyl peroxide.

In the case of sulfur vulcanization, mono- to polysulfur bridges are formed. While the exact mechanism remains unclear, unsaturated precursors of elastomers are needed since the reactive sites are allylic hydrogens (Scheme 2) [6].

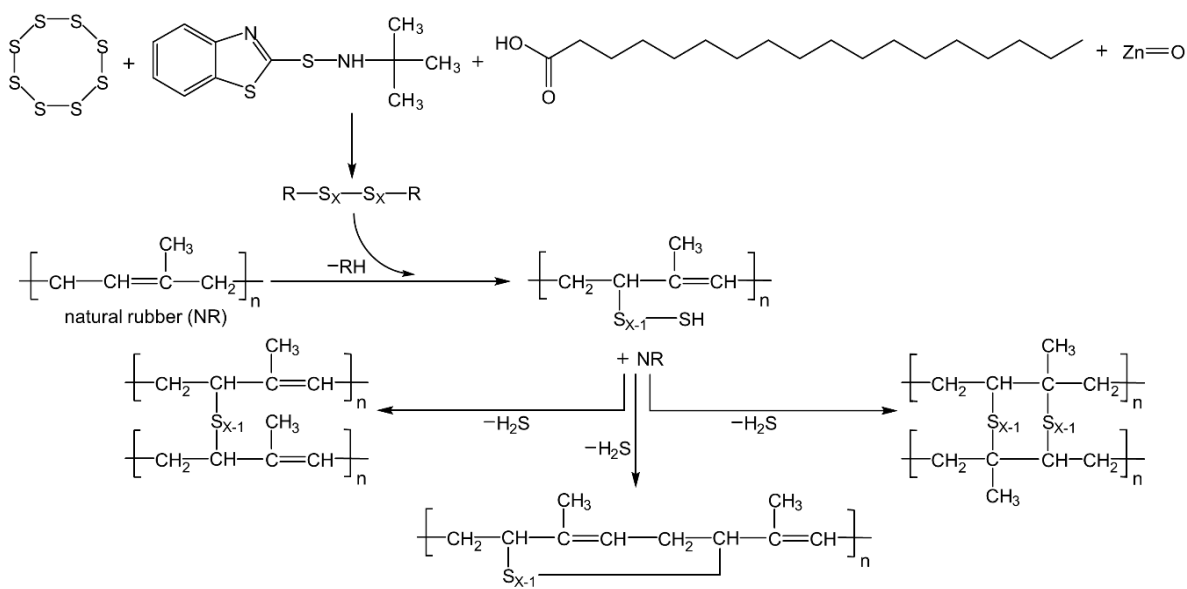

**Scheme 2.** Mechanism of natural rubber cross-linking using sulphur curing.

The method of cross-linking determines the properties of the resulting elastomer. In the case of sulfur cross-linking, the formation of $-S_x-$ polysulfur bonds leads to excellent mechanical properties in terms of flexibility and elongation at break [8]. On the contrary, cross-linking with peroxide agents generates C–C bonds ensuring high mechanical strength, good wear resistance, high thermo-oxidative stability and low compression strength [9].

The valorization of renewable resources in the polymer field currently attracts much attention in academia and industry. Taking into account that an important part of elastomer materials is produced from petrochemical sources, it is desirable to explore the use of novel cross-linked materials of biomass origin, coming from e.g. vegetable oils [10, 11]. Bio-resource has been already successfully implemented for elastomers production. Genencor (now DuPont) with Goodyear Tire & Rubber Company co-developed a process of BioIsoprene[TM] synthesis through the fermentation of glucose using genetically engineered *E.coli* [12] and used such BioIsoprene for the production of synthetic rubber [13].

One good example of low Tg bio-based polymers that can be further cross-linked are estolides. Estolides are aliphatic polyesters that occur in Nature [14], or can be obtained by polycondensation of fatty acids from their hydroxyl or olefinic groups with terminal carboxylic group [15, 16]. This method of polymerization proceeds from AA- and BB-type monomers or AB-type ones [17, 18]. During this step-growth polymerization, low molecular weight by-products, e.g. water or alcohol, are usually formed. In order to shift the equilibrium toward the polymer formation, these by-products have to be eliminated *in situ*. Therefore, the polycondensation reaction usually proceeds in bulk under high temperature, vacuum or gas stream or in the presence of solvent, which forms a heteroazeotropic mixture with the by-product [19]. Usually, estolides are produced at high temperature using strong acid catalysts (perchloric acid, sulphuric acid, *p*-toluenesulfonic acid) [20, 21]. However, the synthesis of estolides can be also performed in milder reaction conditions of temperature and pressure in the presence of lipases [22] or ionic liquids [23].

Estolides found already applications in many areas such as base oils for lubricants and functional fluids [24], viscosity controllers in chocolates, cutting base fluids in metal processing, emulsifiers in margarine and pigment dispersants in paints, inks and cosmetics [25]. Besides renewability, another advantage of estolides is their biodegradability [26].

Estolides can be easily obtained from ricinoleic acid (12-hydroxy-9-*cis*-octadecenoic acid), which is the main constituent of castor oil (85%). It is a very interesting vegetable oil-based monomer, because it contains both hydroxyl and carboxyl group and an additional unsaturated C=C double bond, making it cross-linkable and polymerizable. Additionally, once polymerized, the dangling chain of ricinoleic acid ensures hydrophobicity and acts as an internal plasticizer, having an influence on the physical and mechanical properties of the resulting polyesters [16].

The synthesis of estolides was already performed using lipase catalyst starting from ricinoleic acid [27, 28, 29] or methyl ricinoleate [30, 31]. Ricinoleic acid was also copolymerized with other monomers such as maleic and succinic anhydrides [32], diacid oligomers of poly(sebacic acid) [33], lactic acid [34, 35, 36, 37] or bile acid [38].

Estolides from methyl ricinoleate were already cured by peroxide or sulfur vulcanization to form elastomers with good mechanical performance. However, the cross-linking was performed for the estolides coming from lipase-catalyzed polycondensation process[30, 31]. Moreover, the synthesis of estolide from methyl 12-hydroxystearate and its further vulcanization was not described, to the best of our knowledge.

In this work, methyl ricinoleate (MRic) and methyl 12-hydroxystearate (MHS) were (co)polymerized, to obtain polyester pre-polymers that were subsequently cross-linked to design original bio-based elastomers. Titanium (IV) isopropoxide, Ti(O$i$Pr$_4$), was used as the polycondensation Lewis acid catalyst since titanium-based organometallic catalysts are powerful in the ester-exchange reactions [39]. The polyesters were cross-linked by means of dicumyl peroxide or sulfur vulcanization system. The resulting elastomers were characterized in terms of thermal and mechanical processes.

## 2. Materials and methods

Methyl 12-hydroxystearate (96%) and methyl ricinoleate (93%) were supplied by ITERG company. Methyl 12-hydroxystearate and methyl ricinoleate were purchased from Nu-Chek-Prep Inc. (Elysian, MN, USA, ≥99%). Titanium (IV) isopropoxide (Ti(O$i$Pr$_4$), 99.9%), dicumyl peroxide (DCP, 98%), 2,2'-azobis(2-methylpropionitrile) (AIBN, >98%), dichloromethane and tetrahydrofuran were purchased from Sigma-Aldrich. Benzoyl peroxide (BPO, 75%) and lauroyl peroxide (97%) were purchased from Fischer Scientific. Sulfur curing system i.e. sulfur, zinc oxide, stearic acid and N-tert-butyl-2-benzothiazyl sulfenamide (TBBS) accelerator were supplied by EMAC company. All products and solvents (reagent grade) were used as received.

**Synthesis of estolides from methyl ricinoleate and methyl 12-hydroxystearate**

The polycondensation reactions were carried out in a round bottom flask equipped with a temperature-controlled oil bath and magnetic stirring. Methyl ricinoleate and methyl 12-hydroxystearate were dried at 70°C under vacuum overnight prior to use. Next, each monomer was cooled down to room temperature under vacuum and titanium (IV) isopropoxide (1 wt. %) in dichloromethane was added under nitrogen flow. The mixture was stirred at room temperature during 30 min under nitrogen atmosphere. Afterward, dynamic vacuum was applied. After 30 min of stirring at 70 °C, the temperature was raised to 120 °C for 1 h, subsequently to 140 °C for 1 h and then to 180 °C for around 69 h.

Heating of the reaction mixture under nitrogen atmosphere and then under dynamic vacuum, in the presence of a transesterification catalyst (Ti(O*i*Pr)$_4$), allowed the easy removal of the methanol condensate.

**General procedure for the estolide cross-linking**

Mixtures of the estolides with a curing agent (dicumyl peroxide or sulfur system) were prepared using micro-conical twin screw compounding extruder (Thermo Scientific HAAKE™ MiniCTW) by mixing during 30 min at 120 °C using screw speed of 50 rpm. The obtained blends were cross-linked by compression molding under the hydraulic press at the pressure of 10 bar during 1 h at 160 °C (sulfur vulcanization) or 170 °C (peroxide vulcanization). The sulfur system consisted of 100 phr of the estolide, 3 phr of ZnO, 2 phr of stearic acid, 4 phr of sulfur and 15 phr of TBBS, (where phr means parts per hundred rubber). The amount of dicumyl peroxide applied in the mixture with estolide was equal to 10 wt. %.

**Swelling tests**

Swelling tests were carried out by immersing samples of about 30 mg into 10 ml of THF for 48 h. Samples were then dried under vacuum oven for 48 h at 40 °C. The gel fraction was determined from the equation $GF = \frac{W_d}{W_0} \cdot 100\%$, whereas soluble fraction was determined as $SF = \frac{W_0 - W_d}{W_0} \cdot 100\%$, where $W_0$ - the mass of the sample before the swelling test, $W_d$ - the mass of the dried sample.

**Characterizations**

NMR spectra were recorded using a Bruker AC-400 NMR, at room temperature, in deuterated chloroform ($CDCl_3$).

Fourier Transform Infrared (FT-IR) spectra were performed on a Bruker VERTEX 70 spectrometer equipped with diamond crystal (GladiATR PIKE technologies) for the attenuated total reflection (ATR) mode. The spectra were acquired from 400 to 4000 $cm^{-1}$ at room temperature using 32 scans at a resolution of 4 $cm^{-1}$.

Polymer molar masses were determined by size exclusion chromatography (SEC) using tetrahydrofuran (THF with 250 ppm of BHT as an inhibitor) as an eluent and trichlorobenzene as a flow marker. Measurements in THF were performed on a ThermoFisher Scientific Ultimate 3000 system equipped with Diode Array Detector, Wyatt light scattering detector and RI detector. The separation was achieved on three Tosoh TSK gel columns: G4000HXL (particles of 5 mm, pore size of 200 Å, exclusion limit of 400 000 g $mol^{-1}$), G3000HXL (particles of 5 mm, pore size of 75 Å, exclusion limit of 60 000 g $mol^{-1}$), G2000HXL (particles of 5 mm, pore size of 20 Å, exclusion limit of 10 000 g $mol^{-1}$), at flow rate of 1 mL $min^{-1}$. The injected volume was 20 µL. Columns' temperature was held at 40 °C. Data were recorded and processed by Astra software from Wyatt. SEC was calibrated using polystyrene standards.

Thermogravimetric analysis (TGA) thermograms were obtained using a TGA Q500 apparatus from TA instruments. Samples (~10 mg) were heated from room temperature to 700 °C at a rate of 10 °C $min^{-1}$ under nitrogen atmosphere.

Differential scanning calorimetry (DSC) measurements of samples (~5 mg) were performed using a DSC Q 100 apparatus from TA Instruments over temperature range from -120 °C to 240 °C, in a heating-cooling mode of 10 °C $min^{-1}$. The analyses were carried out in a nitrogen atmosphere using aluminum pans. Glass transition temperatures and melting points were obtained from the second heating runs.

Dynamic mechanical analysis (DMA) measurements were carried out using a RSA3 apparatus from TA Instruments equipped with a liquid nitrogen cooling system. The thermomechanical properties of the samples (width ≅ 5 mm; thickness ≅ 2 mm and length of the fixed section ≅ 10 mm) were studied from around -70 °C to $T_m$ + 15 °C at a heating rate of 3 °C min$^{-1}$. The measurements were performed in tensile test at a frequency of 1 Hz, a strain sweep of 0.03% and an initial static force of 0.1 N.

Tensile tests were performed on extension mode test system using a RSA3 apparatus from TA Instruments. The measurements were carried out at room temperature for rectangular samples (width ≅ 5 mm; thickness ≅ 2 mm and length of fixed section ≅ 10 mm) using a crosshead speed of 2 mm min$^{-1}$.

## 3. Results and discussion

**Part 1: Pre-polymer synthesis and characterization**

Methyl ricinoleate (MRic) and methyl 12-hydroxystearate (MHS) were transesterified using Ti(O*i*Pr)$_4$ as a catalyst to prepare three estolides: polyricinoleate (PRic), polyhydroxystearate (PHS) and an equimolar copolymer of methyl ricinoleate and methyl 12-hydroxystearate P(Ric+HS). The polymer structures were confirmed by $^1$H NMR and FT-IR spectroscopies (Fig. 1 and Fig. S1–S5).

It can be observed, that in all $^1$H NMR spectra, the characteristic peak of the methine proton –C**H**– (c) at 3.59 ppm linked to the carbon bearing the hydroxyl function in methyl ricinoleate and methyl 12-hydroxystearate is shifted downfield to 4.86 ppm (m) confirming the formation of ester linkages. Moreover, the lower intensity of the methyl protons –OC**H**$_3$ (k) proves the reaction of methyl ester functions with hydroxyl functions of methyl ricinoleate or methyl 12-hydroxystearate. Finally, by FT-IR spectroscopy the bands coming from the vibrations of hydroxyl groups in the wavenumber range between 3000 cm$^{-1}$ and 3500 cm$^{-1}$ disappear as expected (see Figs. S3–S5).

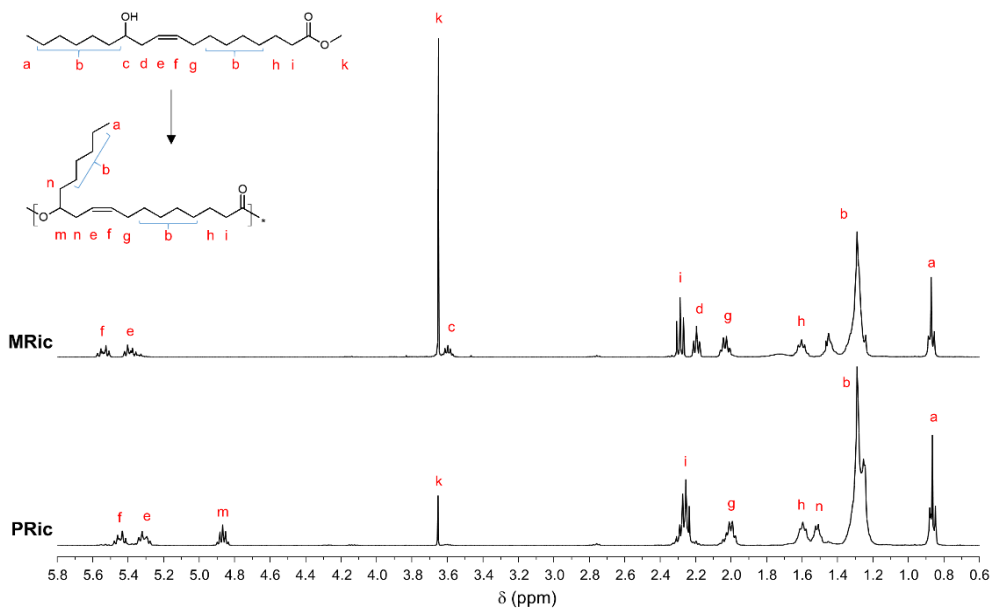

**Fig. 1.** $^1$H NMR of methyl ricinoleate (MRic) and polyricinoleate (PRic).

The synthesis of estolides was also monitored using SEC. Fig. 2 and 3 show SEC traces of the reaction media during polycondensation of methyl ricinoleate (see Supporting Info Fig. S6–S9 for the traces collected during the synthesis of PHS and P(Ric + HS). Fig. 2 presents SEC traces of PRic A from monomers with a GC purity ≥ 99% (Nu-Chek-Prep Inc.) whereas Fig. 3 shows SEC traces of PRic B from monomers with a GC purity of 93%, as received from ITERG. The shift of the SEC traces to lower elution times with reaction progress revealed the efficiency of the AB self-polycondensation. The values of $M_w$, weight average molecular weight relative to PS calibration, and Đ, dispersity, obtained after 8 h of transesterification are noted in Table 1. PRic A exhibited the molecular weight $\overline{M_n}$ in the range of ~7 kg mol$^{-1}$ and $\overline{M_w}$ of 19–22 kg mol$^{-1}$, while PRic B exhibited lower values of 3–4 kg mol$^{-1}$ and 5–8 kg mol$^{-1}$, respectively. The dispersity values were in the range 1.8–3.1. In the case of estolides (PRic B) obtained from less pure monomers, oligomers persist with time (Fig. S8-S9). One can explain this feature by the presence of saturated fatty acid that could act as chain stopper.

The increase of the polycondensation time allowed enhancement of the molecular weight of estolides up to $\overline{M_w}$ values of 62.6 kg mol$^{-1}$ (Table 1, Fig. S10). However, dispersity increased dramatically (Đ>4) indicating the occurrence of side reactions such as etherification. [40]

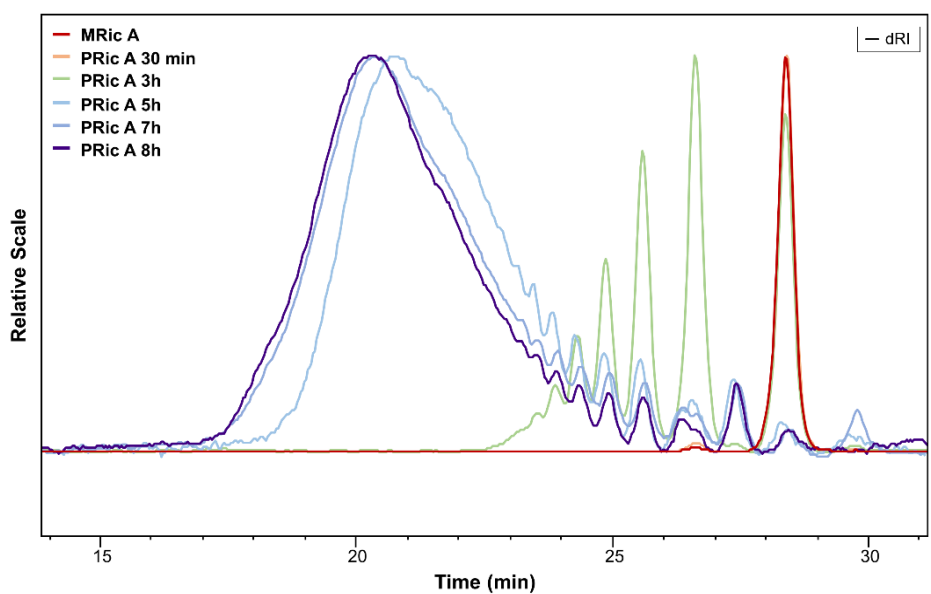

**Fig. 2.** SEC traces of methyl ricinoleate (MRic A) and PRic A estolides.

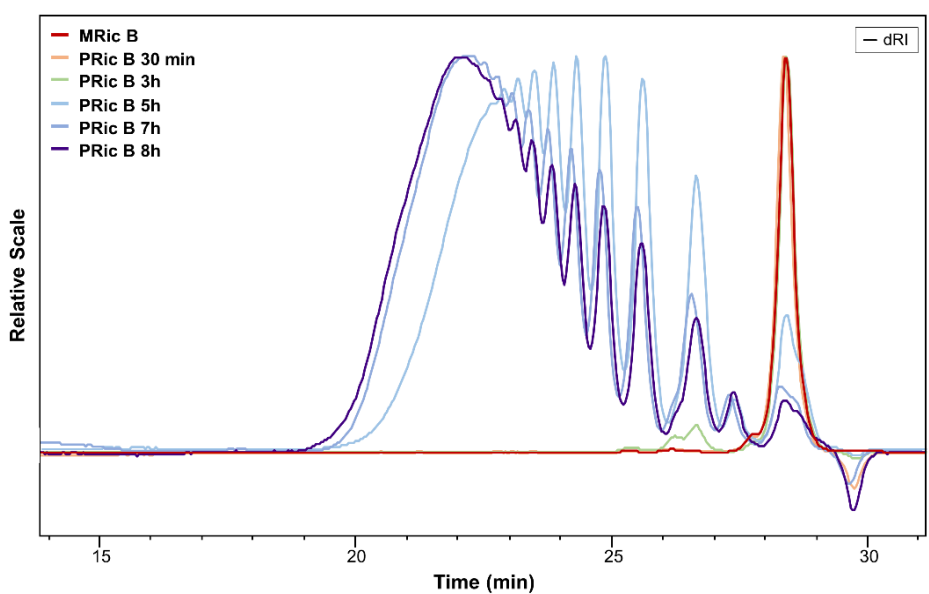

**Fig. 3.** SEC traces of methyl ricinoleate (MRic B) and PRic B estolides.

**Table 1.** Molecular characteristics of synthesized estolides.

| Estolide | Monomer purity (%) | $\overline{M_w}$ (kg mol$^{-1}$) | Đ |
|---|---|---|---|
| PRic A 8 h | 99 | 21.9 | 3.1 |
| PHS A 8 h | 99 | 19.4 | 3.0 |
| P(Ric + HS) A 8 h | 99 | 22.3 | 3.1 |
| PRic A 72 h | 99 | 59.2 | 4.5 |
| PHS A 72 h | 99 | 33.4 | 4.9 |
| P(Ric + HS) A 72 h | 99 | 62.6 | 10.1 |
| PRic B 8 h | 93 | 5.4 | 1.7 |
| PHS B 8 h | 96 | 8.4 | 2.2 |
| P(Ric + HS) B 8 h | 94.5 | 5.1 | 1.8 |
| PRic B 72 h | 93 | 11.5 | 1.5 |

The good thermal stability of the estolides was confirmed by TGA (Table 2, Fig. S11). Temperatures of the 5% of degradation are high and their values are in the range of 306–317 °C. Polyricinoleates (PRic A 72h, PRic B 72 h) exhibit glass transition temperatures of -68 °C and -74 °C, respectively (Fig. 4). The 9°C difference is due to the lower molecular weight of PRic B and the presence of oligomers, which act as plasticizers. While PRics are fully amorphous, polyhydroxystearate (PHS A) is a semi-crystalline polymer with a melting point of -48°C ($\Delta H_f = 33\,J/g$). Finally the copolymer of methyl ricinoleate and methyl 12-hydroxystearate (P(Ric + HS) A 72 h) shows an intermediate behavior with a $T_g$ of -63°C and a melting temperature of -22°C ($\Delta H_f = 3\,J/g$).

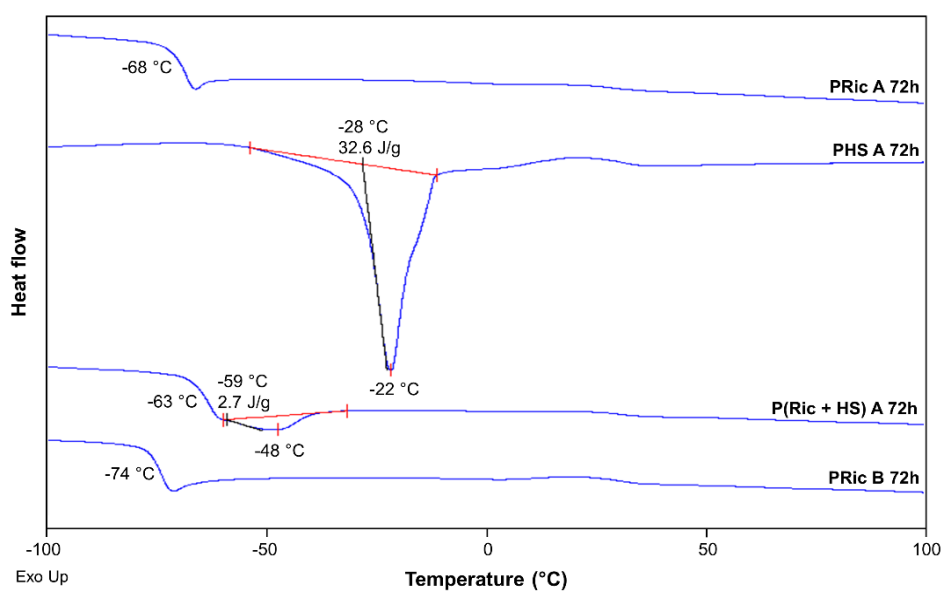

**Fig. 4.** DSC traces of the estolides obtained after 72 h polycondensation.

**Table 2.** Data of thermal analyses of high molecular weight estolides (PRic A 72 h, PHS A 72 h, P(Ric + HS) A 72 h, PRic B 72 h). Values of char at 500 °C and temperatures corresponding to 5% degradation, glass transition and melting point of estolides.

| Estolide | $T_{5\%}$ (°C) | Char at 500°C (%) | $T_g$ (°C) | $T_m$ (°C) |
|---|---|---|---|---|
| PRic A 72 h | 315 | 0.9 | -68 | – |
| PHS A 72 h | 311 | 2.4 | – | -22 |
| P(Ric + HS) A 72 h | 317 | 1.2 | -63 | -48 |
| PRic B 72 h | 306 | 0.8 | -74 | – |

One fascinating property of estolides, especially polyricinoleate, which contains double bonds, is their ability to cross-link. A cross-linked material can be created by various processes such as high-temperature treatment, irradiation or by adding cross-linking agents [41]. However, the cross-linking usually proceeds using sulfur or peroxide vulcanization. Because of the absence of unsaturation, polyhydroxystearate cannot be vulcanized with sulfur and is not very reactive towards peroxide curing. On the other hand, polyricinoleate and copolymer of methyl ricinoleate and methyl 12-hydroxystearate contain unsaturated double bonds in their structures. Thus it is possible to obtain cross-linked materials using both sulfur and peroxide vulcanization.

**Part 2: Synthesis of elastomers from estolides**

First, polyricinoleate cross-linking ability was determined by DSC on-line curing of PRic B 72 h (Fig. 5) with sulfur system or common peroxides such as 2,2'-azobis(2-methylpropionitrile) (AIBN), benzoyl peroxide (BPO), dicumyl peroxide (DCP) and lauroyl peroxide (LPO).

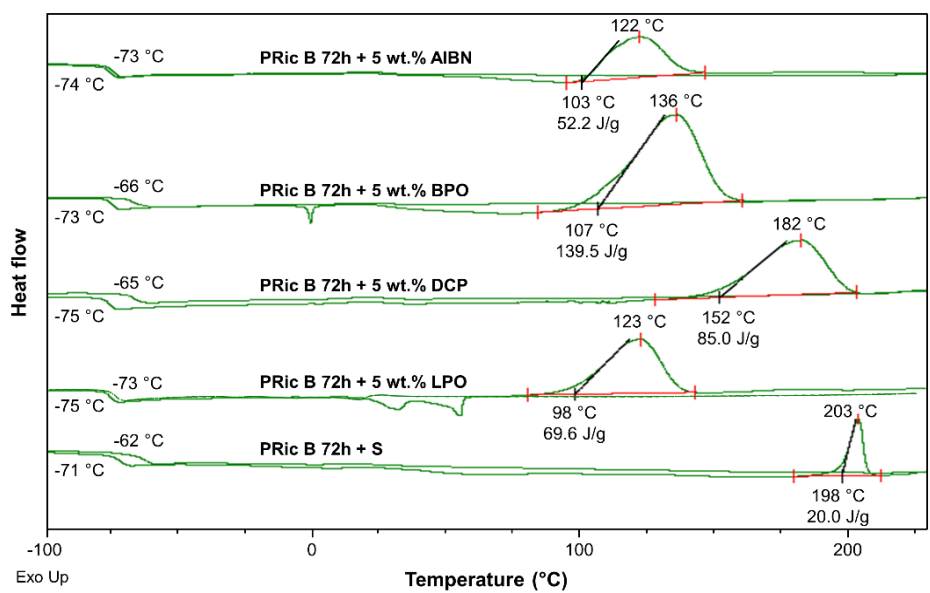

**Fig. 5.** DSC on-line curing of PRic B 72 h with different peroxides and sulfur system. (heating-cooling rate of 10 °C min$^{-1}$)

Homogeneous mixture of the estolide and the curing agent is critical to obtain a uniform network. To ensure good mixing, the viscosity of the estolides need to be decreased by working at high temperature, i.e. 120 °C. Thus the curing needs to occur at a higher temperature. Based on this statement and from DSC on-line curing, sulfur and dicumyl peroxide systems were selected for further study. The cross-linking tests were carried out using compression molding at 170 °C (peroxide vulcanization) or 160 °C (sulfur vulcanization) for 1 h after vigorous mixing of the components by extruder at 120 °C.

The formation of the network (cross-linking) was confirmed by THF swelling tests determining the soluble fraction (SF) and the gel fraction (GF) (Table 3). As expected, the gel fraction values were higher in the case of PRic A 72 h pre-polymer in regard to PRic B 72 h. Besides, one could observe the higher efficiency of DCP (entries 1, 4, 6 in Table 3) in comparison to sulfur (entries 2, 5, 7 in Table 3). The GF values of the cross-linked estolides after solubility tests were higher in the case of pre-polymers cured using DCP. In the case of sulfur vulcanization of estolides, GF values were lower since SF consists mainly from sulfur vulcanization system, which did not take part in the cross-linking process. Moreover, it can be observed that the SF decreases with increasing $\overline{M_w}$ value of the estolide. The SF value of DCP-cross-linked PRic A 72 h (59.2 kg mol$^{-1}$) is equal to 10% while 13% are obtained for PRic B 72 h (11.5 kg mol$^{-1}$). Same trend is observed for sulfur-cross-linked PRic(i.e. the SF values are 18% and 24%, respectively). The decrease of SF with the molecular weight can be explained by the increasing probability of cross-linking of estolide having higher molecular weight.

The soluble fractions (SF) were analyzed by SEC chromatography (Table 3, Fig. S12) and [1]H NMR (Fig S13). The latter appeared to be mainly constituted of short chains with a decrease of $\overline{M_w}$ by a factor 2–20 compared to the initial polymer. Moreover in the case of sulfur cross-linking, TBBS is also a non-neglectible part of the SF.

Estolide PHS A 72 h cross-linked using DCP possess the highest GF value of 96% among analyzed cross-linked materials. PHS A 72 h having $\overline{M_w}$ value of 33.4 kg mol$^{-1}$ didn't contain C=C double bonds in the structure causing the radical cross-linking process only through hydrogen abstraction mechanism, followed by combination of the resulting macroradicals forming very stable C–C single bonds. [42].

**Table 3.** Solubility test results of cross-linked estolides. Molecular weight distribution of the soluble fraction was analyzed by SEC in THF.

| Cross-linked estolide | GF (%) | SF (%) | $\overline{M_w}$ of SF (kg mol$^{-1}$) | Đ of SF |
|---|---|---|---|---|
| PRic A 72 h + DCP | 90 | 10 | 3.0 | 2.5 |
| PRic A 72 h + S | 82 | 18 | 5.1 | 2.3 |
| PHS A 72 h + DCP | 96 | 4 | 5.3 | 2.8 |
| P(Ric + HS) A 72 h + DCP | 95 | 5 | 7.6 | 5.0 |
| P(Ric + HS) A 72 h + S | 84 | 16 | 10.0 | 3.1 |
| PRic B 72 h + DCP | 87 | 13 | 5.4 | 1.9 |
| PRic B 72 h + S | 76 | 24 | 6.7 | 2.5 |

The thermal and mechanical properties of the cross-linked estolides were investigated by TGA (Fig. 6 and S14), DSC (Fig. 7), DMA (Fig. 8, 9) and tensile tests (Fig. 10).

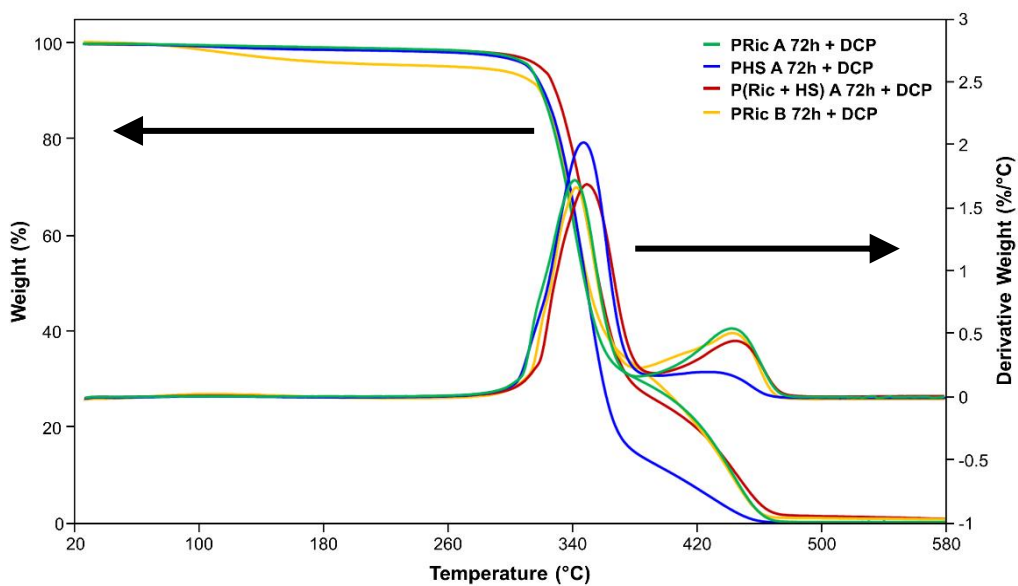

**Fig. 6.** TGA and DTG curves of high molecular weight estolides cross-linked using DCP (PRic A 72 h + DCP, PHS A 72 h + DCP, P(Ric + HS) A 72 h + DCP, PRic B 72 h + DCP).

TGA analysis revealed high thermal stability of estolide networks (Table 4) with, in all cases, a two-step degradation exhibiting maximum around 350°C and 450°C respectively. Higher $T_{5\%}$ values (260–318 °C) were observed for DCP-vulcanized samples in comparison to the ones obtained for sulfur-vulcanized networks ($T_{5\%}$ = 203–217 °C). Char values at 500 °C slightly differed with respect to the method of vulcanization: for DCP cross-linking, these values were significantly lower (0.6–1.5%) in comparison to the sulfur cross-linking method (4.1–7.7%).

DSC was employed to gain insight into the evaluation of glass transition temperature being one of the fundamental indicators of the cross-linking effect (Table S1). $T_g$ values of the estolide networks were ranging from -69 °C to -54 °C. These values were slightly higher of 4–7 °C (DCP vulcanization) and 9–14 °C (sulfur vulcanization) to the ones of estolide precursors. Surprisingly, cross-linked PHS exhibits a melting point ($T_m$ = -24 °C) as its precursor with a slight decrease of the melting enthalpy by a factor of 1.6. This crystallization may be due to the 4 % of the soluble uncross-linked polymers.

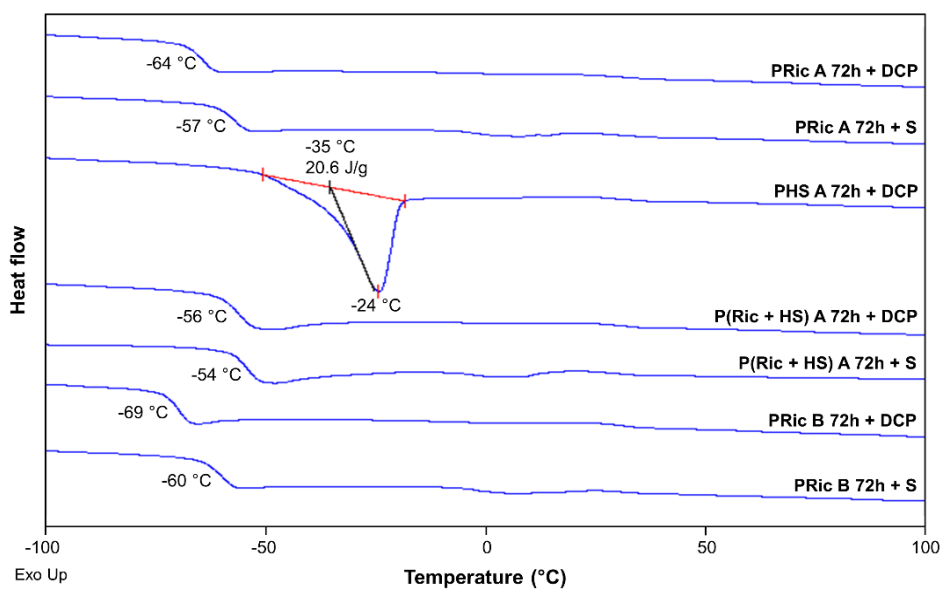

**Fig. 7.** DSC analysis of cross-linked estolides (PRic A 72 h + DCP, PRic A 72 h + S, PHS A 72 h + DCP, P(Ric + HS) A 72 h + DCP, P(Ric+HS) A 72 h + S, PRic B 72 h + DCP, PRic B 72 h + S).

While cross-linked PHS was too brittle to be further analyzed, thermomechanical properties of PRic and P(Ric + HS) cross-linked estolides were analyzed by DMA (Fig. 8, 9). The alpha transition temperature ($T_\alpha$) being a mechanical equivalent of the glass transition temperature ($T_g$) and corresponding to the transition from a glassy to a rubbery state, was determined. It relates to the maximum of the tan δ curve. Moreover, the density of cross-linking or the effective number of network junctions per unit volume ($v$) was calculated from the following equation according to rubber-like elasticity theory:

$$v = \frac{E'_{T_{\alpha+30\,K}}}{\phi \cdot R \cdot T_{\alpha+30\,K}}$$

where E' is the storage modulus of the network at $T_{\alpha\,+\,30\,K}$, $T_\alpha$ is the maximum of tan δ, φ is the front factor approximately equal to 1 in Flory theory and R = 8.314 J mol$^{-1}$ K$^{-1}$ is the gas constant.

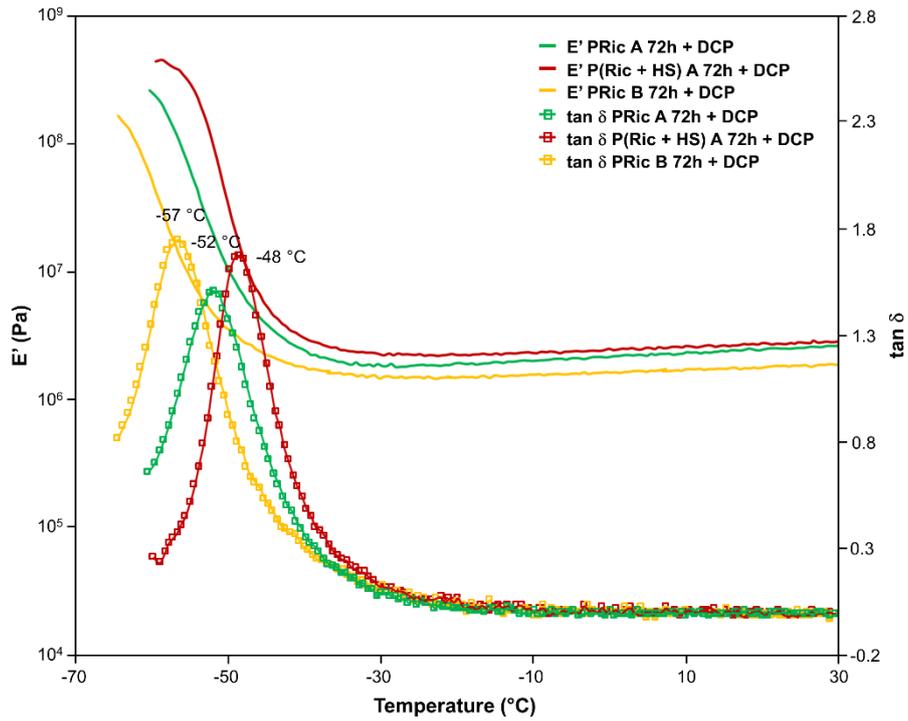

**Fig. 8.** DMA analysis of estolides cross-linked using DCP: PRic A 72 h + DCP, P(Ric + HS) A 72 h + DCP, PRic B 72 h + DCP.

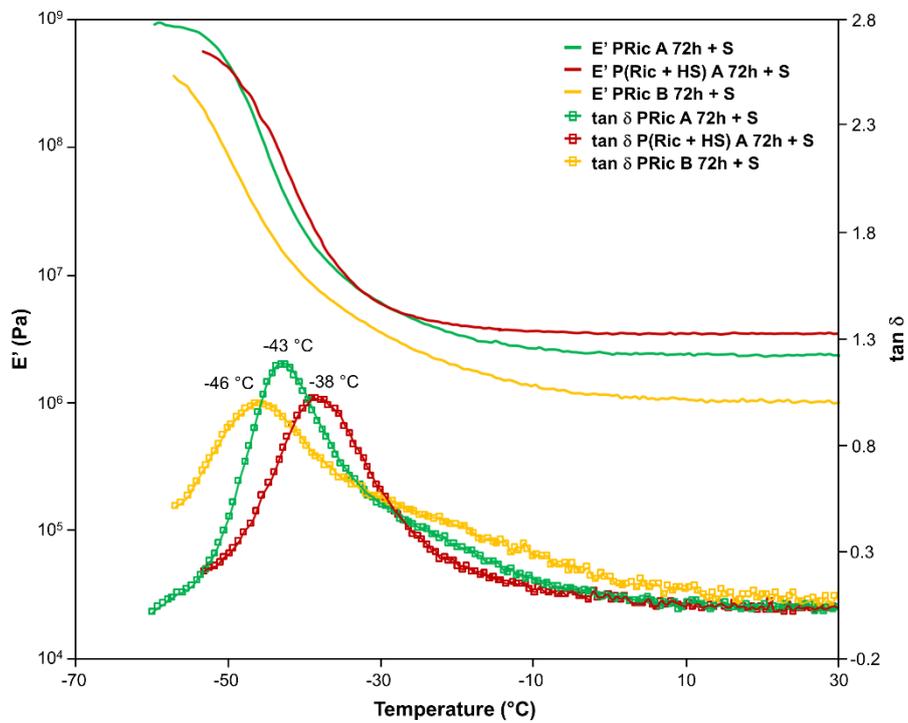

**Fig. 9.** DMA analysis of estolides cross-linked using sulfur system: PRic A 72 h + S, P(Ric + HS) A 72 h + S, PRic B 72 h + S.

Figs. 8 and 9 show a plot overlay of storage modulus E' and loss factor tan δ of cross-linked high molecular weight estolides (PRic A 72 h + DCP, PRic A 72 h + S, PHS A 72 h + DCP, P(Ric + HS) A 72 h + DCP, P(Ric+HS) A 72 h + S, PRic B 72 h + DCP, PRic B 72 h + S) as a function of temperature. All the thermomechanical data of cross-linked estolides are given in Table 4.

In all cases, DMA curves are typical of cross-linked materials with two plateaus, glassy zone and rubbery plateau respectively before and after the $T_\alpha$. Cross-linked estolides by DCP exhibit slightly lower glassy modulus than sulfur cross-linked estolides. Besides, as expected this plateau is lower for Pric B whatever the cross-linking systems used. $T_\alpha$ values are in agreement to the Tg obtained by DSC with a slight shift toward higher temperature. Moreover, one can observe different widths of the tan δ peaks. Narrower tan δ peaks are obtained for DCP-cross-linked estolides whereas broader ones are noticed for sulfur-cross-linked estolides. A broad tan δ peak is usually due to inhomogeneity in the network, which in the case of sulfur systems, could be explained by the nature of $-S_x-$ bridges. Indeed a -S- bridge is completely different, in term of mobility for example, from a $-S_8-$ bridge. Finally, the rubbery plateau confirms the cross-linking of the materials and its value is directly proportional to the cross-linking density (Table 4). Contrary to the glassy zone, this plateau is slightly higher in the case of sulfur cross-linking and is lower for Pric B in both cases. Surprisingly the highest plateau is obtained for the copolymer P(Ric+HS), this may be explained by the lower SF for this copolymers compared to Pric A (16% compared to 18% for the sulfur system). Indeed the soluble fraction will act as a plasticizer decreasing the storage moduli of the materials.

**Table 4.** Thermomechanical properties of cross-linked estolides: PRic A 72 h + DCP, PRic A 72 h + S, P(Ric + HS) A 72 h + DCP, P(Ric+HS) A 72 h + S, PRic B 72 h + DCP, PRic B 72 h + S.

| Cross-linked estolide | $\tan \delta$ | $T_\alpha$ (°C) | $E'_{T_\alpha+30\,K}$ (MPa) | $\nu$ ($10^3$ mol m$^{-3}$) |
|---|---|---|---|---|
| PRic A 72 h + DCP | 1.50 | -52 | 1.8 | 0.87 |
| PRic A 72 h + S | 1.18 | -43 | 2.8 | 1.29 |
| P(Ric + HS) A 72 h + DCP | 1.67 | -48 | 2.2 | 1.04 |
| P(Ric + HS) A 72 h + S | 1.02 | -38 | 3.5 | 1.59 |
| PRic B 72 h + DCP | 1.75 | -57 | 1.5 | 0.73 |
| PRic B 72 h + S | 0.99 | -46 | 1.7 | 0.80 |

Additionally, tensile tests were performed in order to determine the values of tensile strength and elongation at break of cross-linked estolides (Fig. 10). It can be observed that elastomers obtained from sulfur vulcanization i.e. PRic A 72 h + S and PRic B 72 h + S exhibit lower values of tensile strength (0.11–0.24 MPa) and higher values of elongation at break (25–43%) than elastomers obtained by peroxide vulcanization i.e. PRic A 72 h + DCP and PRic B 72 h + DCP (0.19–0.40 MPa and 13–15%, respectively). Additionally, cross-linked estolides with lower $\overline{M_w}$ of 11.5 kg mol$^{-1}$ (PRic B 72 h) experienced lower elongations at break in comparison to the ones having higher $\overline{M_w}$ of 59.2 kg mol$^{-1}$. Based on the Young modulus values, it can be stated that estolides cross-linked using dicumyl peroxide i.e. PRic A 72 h + DCP and PRic B 72 h + DCP are more stiffen (0.017–0.035 MPa) than estolides cross-linked by sulfur vulcanization system i.e. PRic A 72 h + S and PRic B 72 h + S (0.007–0.010 MPa). The results obtained by tensile tests can be correlated with the material cross-link density. The elongation at break and the tensile strength increase with the increasing cross-link density. The peroxide cross-linked estolides i.e. PRic A 72 h + DCP and PRic B 72 h + DCP with cross-link density values of 0.87 and 0.73 × 10³ mol m$^{-3}$ exhibited worse tensile properties than sulfur-cross-linked ones having values of 1.29 and 0.80 × 10³ mol m$^{-3}$, respectively.

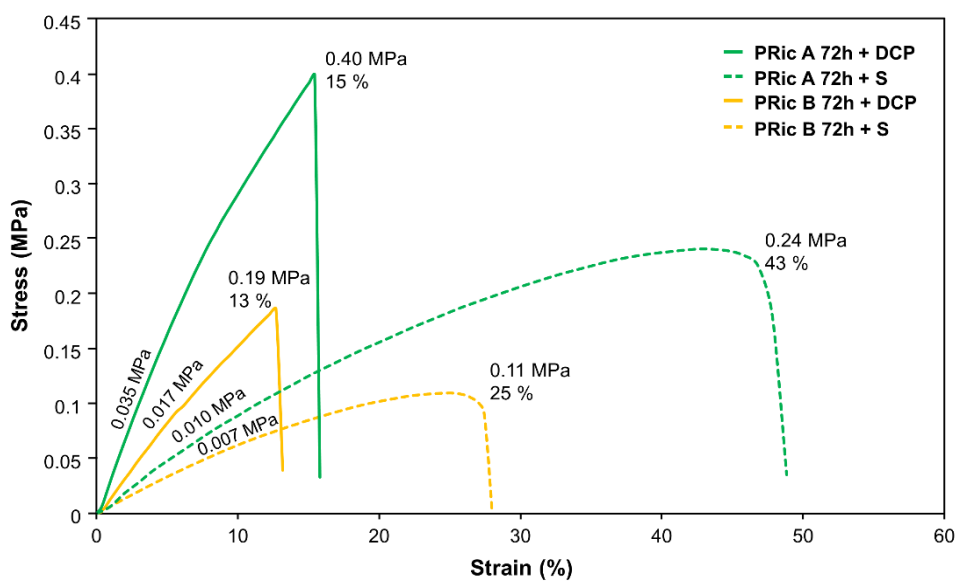

**Fig. 10.** Tensile strength and elongation at break of estolides cross-linked using DCP or sulfur system: PRic A 72 h + DCP, PRic A 72 h + S, PRic B 72 h + DCP, PRic B 72 h + S.

The results of the performed experiments suggest that high molecular weight estolides coming from fatty acid methyl esters such as methyl ricinoleate and methyl 12-hydroxystearate are promising bio-based source of cross-linked materials. So far the data presented in the literature considering the synthesis of high $\overline{M_w}$ estolides (50 – 100 kg mol$^{-1}$) were based on the methyl ricinoleate polycondensation using lipase-catalyzed polycondensation [30, 31]. In this study, we have investigated the polycondensation of methyl ricinoleate (from two different sources: pure commercial one and less pure industrial grade from ITERG company) as well as methyl 12-hydroxystearate and copolymers thereof using titanium (IV) isopropoxide catalyst (Ti(O$i$Pr$_4$)). We were able to obtain high molecular weight estolides with different $\overline{M_w}$ values depending on the monomer purity, ranging from 11.5 kg mol$^{-1}$ to 62.6 kg mol$^{-1}$. The control over the cross-linking of estolides (using dicumyl peroxide or sulfur system) has an influence over the properties of resulted polymers.

## 4. Conclusions

In the present work, methyl esters of fatty acids bearing hydroxyl functions enable esterification to produce estolides. Polyesterification appeared to be a promising methodology for achieving the polymers able to form networks by peroxide or sulfur vulcanization. Elastomers stemming from polyricinoleate, polyhydroxystearate and copolymers thereof were obtained using two different curing agents i.e. dicumyl peroxide and sulfur system. In the case of estolides from methyl 12-hydroxystearate, due to the lack of C=C double bonds, cross-linking was possible only using peroxide curing agent. However, the material was brittle and unable to be analysed by DMA and tensile test. On the other hand, the cross-linked estolides (coming from methyl ricinoleate and copolymer of methyl ricinoleate and methyl 12-hydroxystearate) both exhibited tailored thermal and mechanical properties. Based on the achieved results, they could potentially find applications as replacement of synthetic rubbers.

**Acknowledgments**

The authors gratefully acknowledge the financial support from ADEME (French Environment and Energy Management Agency). The authors sincerely thank the companies for providing the samples of the chemicals: ITERG (French Institute for Fats and Oils) for methyl ricinoleate and methyl 12-hydroxystearate, and EMAC for the sulfur vulcanization system.

Supplementary Information

# Cross-linking of polyesters based on fatty acids

**Sylwia Dworakowska[1,2], Cédric Le Coz[2], Guillaume Chollet[3], Etienne Grau*,[2], Henri Cramail*,[2]**

[1] Cracow University of Technology, Faculty of Chemical Engineering and Technology, Warszawska 24, 31-155 Cracow, Poland

[2] Univ. Bordeaux, CNRS, Bordeaux INP, LCPO, UMR 5629, F-33600, Pessac, France

[3] ITERG, Lipochimie Hall Industriel, 11 rue Gaspard Monge, 33600 Pessac Cedex, France

*Correspondence: (Etienne Grau, Henri Cramail, Laboratoire de Chimie des Polymères Organiques, Université de Bordeaux, UMR5629, CNRS - Bordeaux INP - ENSCBP, 16 Avenue Pey-Berland, 33607 Pessac Cedex, France ; **Fax:** +33 05 40 00 84 87 **Email:** egrau@enscbp.fr, cramail@enscbp.fr)

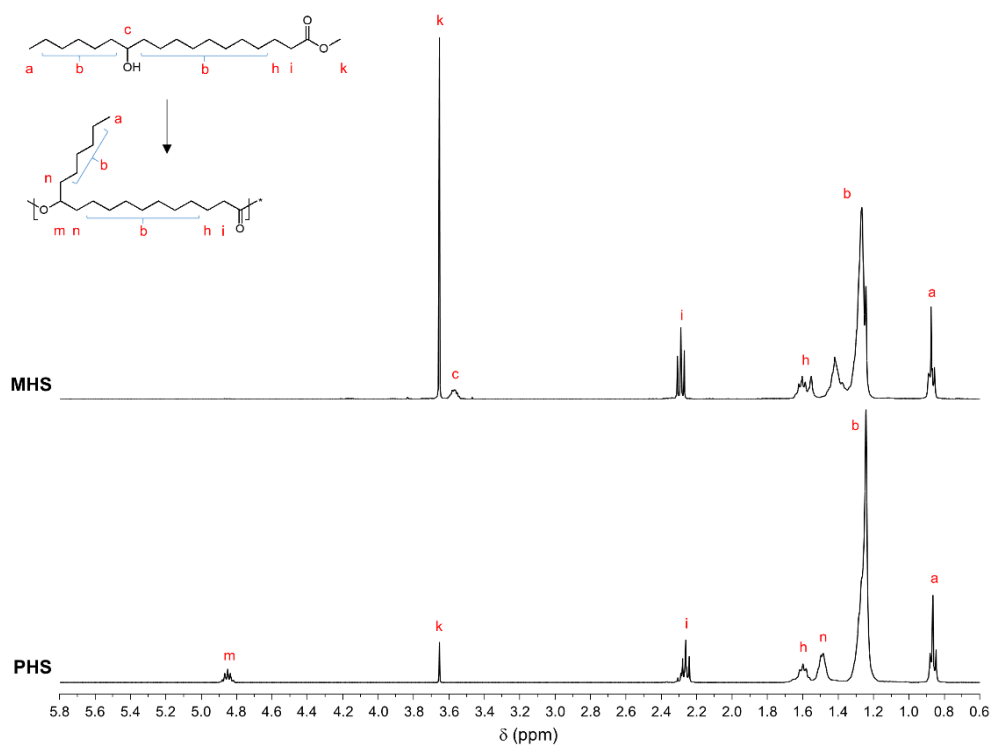

**Fig. S1.** $^1$H NMR of methyl 12-hydroxystearate (MHS) and polyhydroxystearate (PHS).

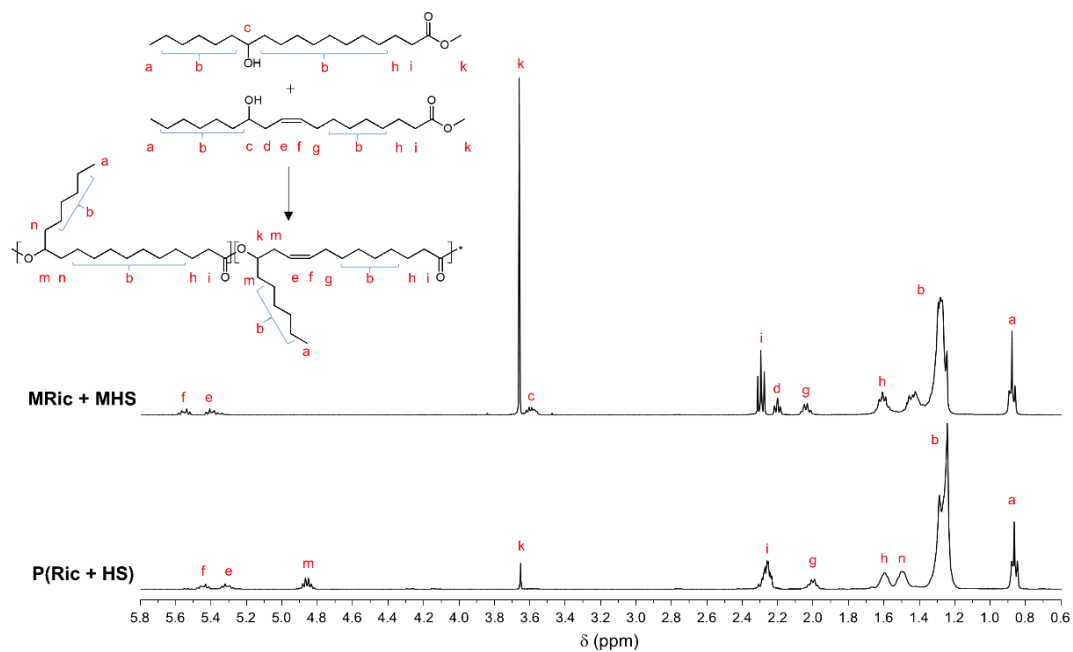

**Fig. S2.** $^1$H NMR of the mixture of methyl ricinoleate and methyl 12-hydroxystearate (MRic + MHS) and copolymer of methyl ricinoleate and methyl 12-hydroxystearate P(Ric + HS).

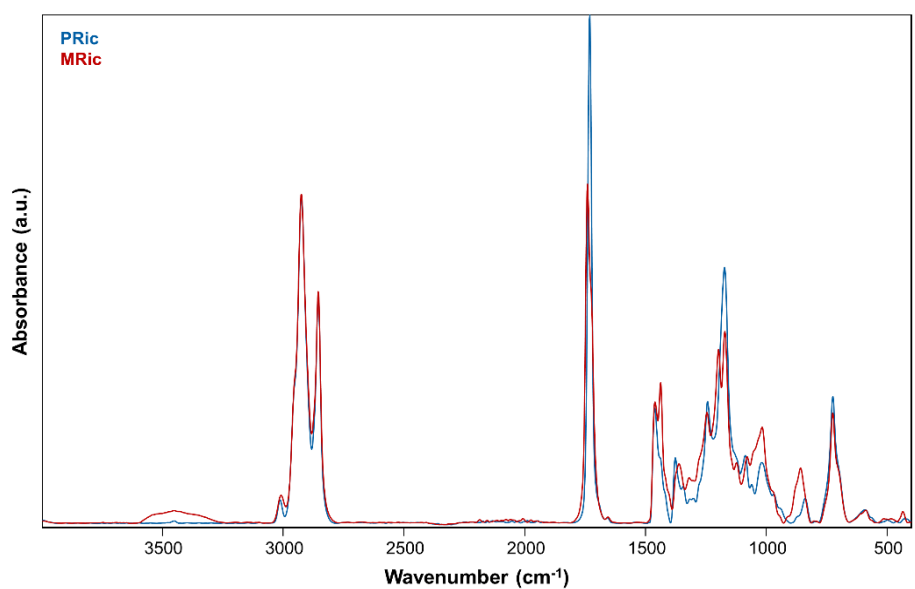

**Fig. S3.** FT-IR spectra of methyl ricinoleate (MRic) and polyricinoleate (PRic).

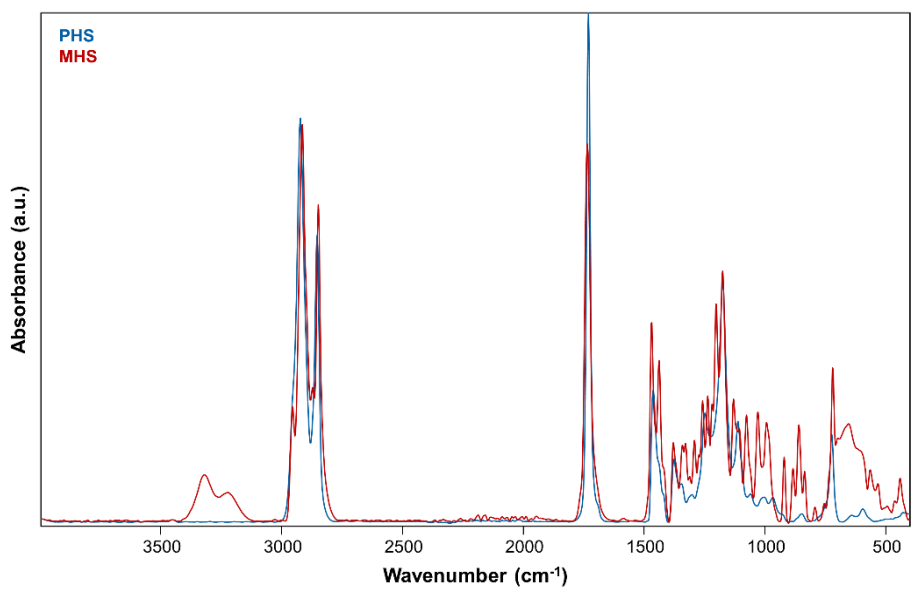

**Fig. S4.** FT-IR spectra of methyl 12-hydroxystearate (MHS) and polyhydroxystearate (PHS).

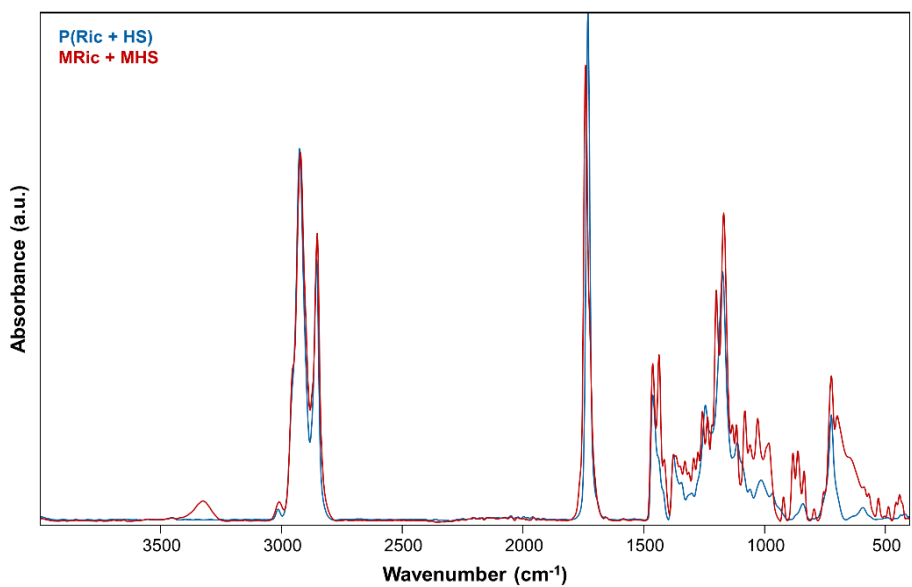

**Fig. S5.** FT-IR spectra of the mixture of methyl ricinoleate and methyl 12-hydroxystearate (MRic + MHS) and copolymer of methyl ricinoleate and methyl 12-hydroxystearate P(Ric + HS).

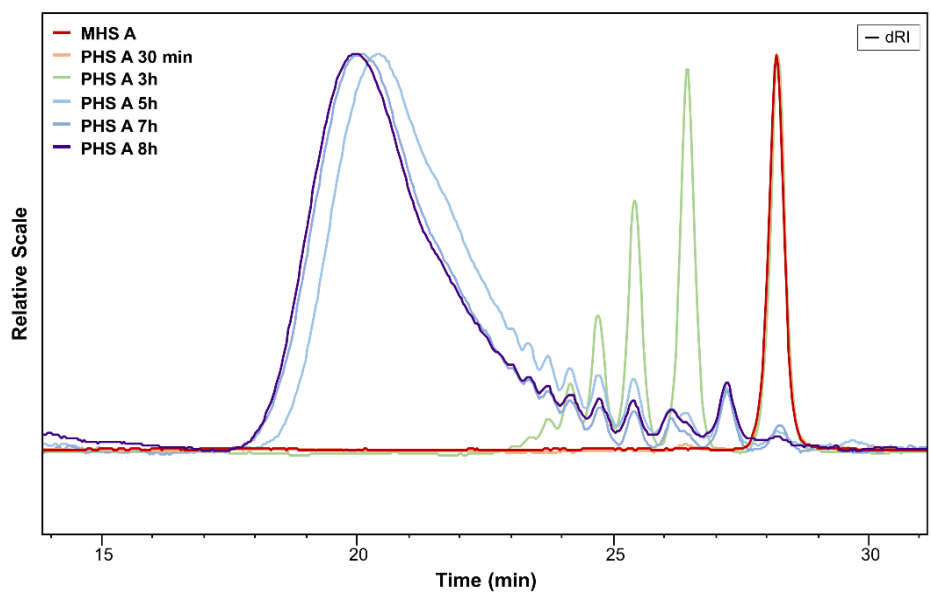

**Fig. S6.** SEC traces of methyl 12-hydroxystearate (MHS A) and PHS A estolides.

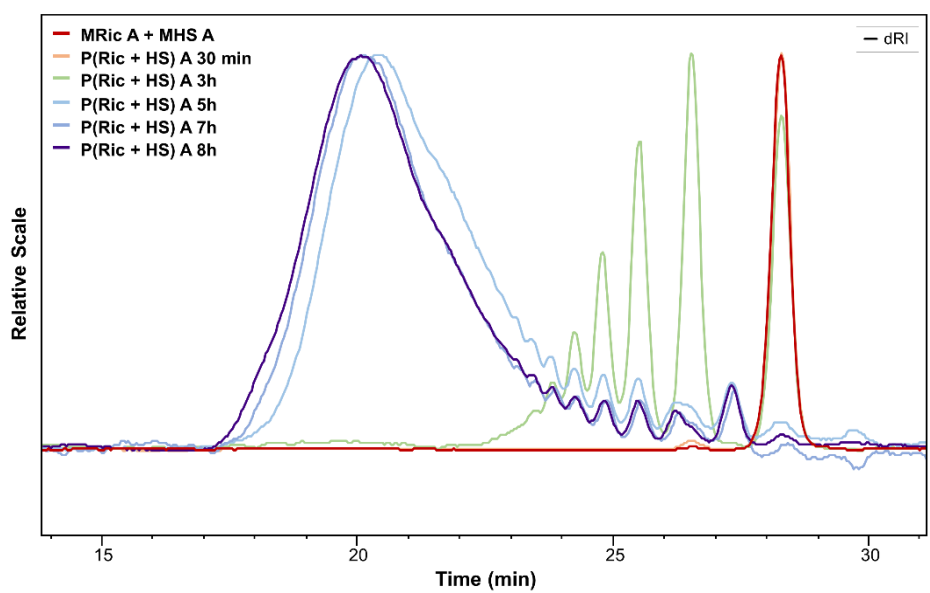

**Fig. S7.** SEC traces of the mixture of methyl ricinoleate and methyl 12-hydroxystearate (MRic A + MHS A) and P(Ric+HS) A estolides.

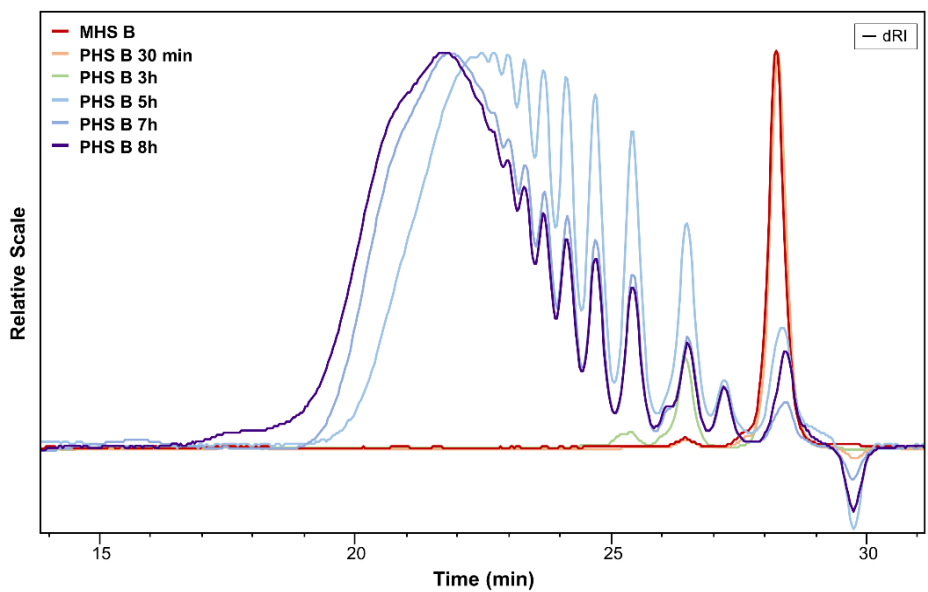

**Fig. S8.** SEC traces of methyl 12-hydroxystearate (MHS B) and PHS B estolides.

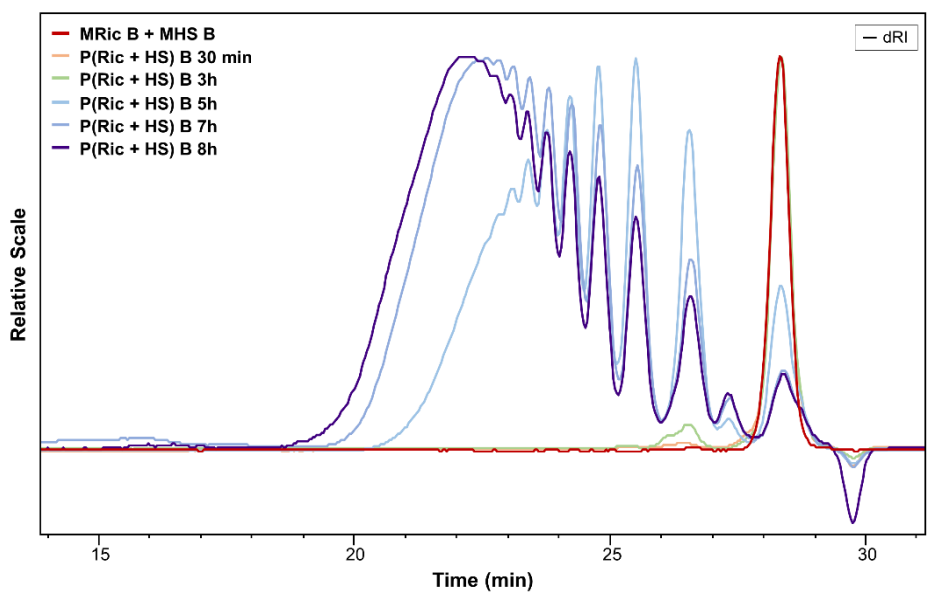

**Fig. S9.** SEC traces of the mixture of methyl ricinoleate and methyl 12-hydroxystearate (MRic B + MHS B) and P(Ric + HS) B estolides.

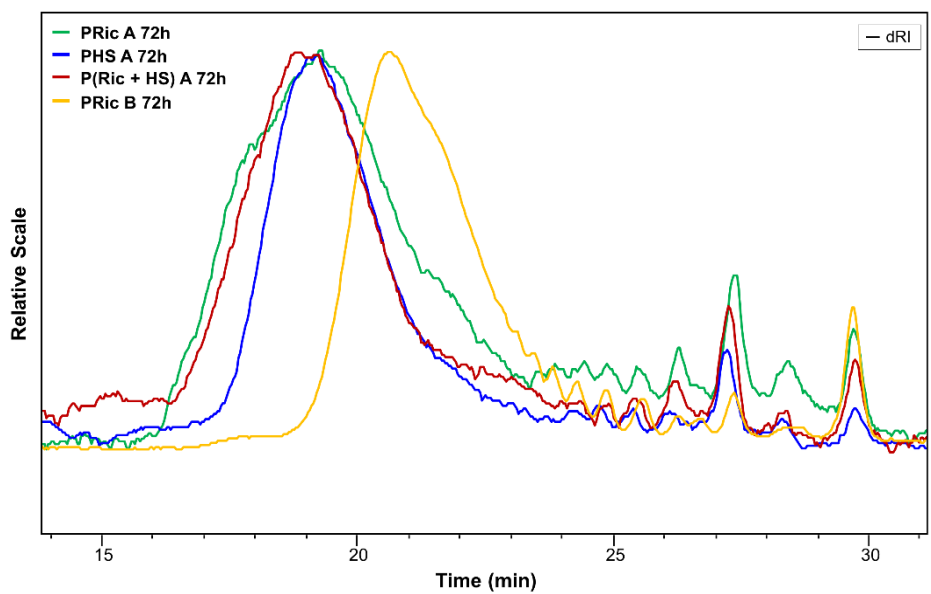

**Fig. S10.** SEC traces of the high molecular weight estolides: PRic A 72 h, PHS A 72 h, P(Ric + HS) A 72 h, PRic B 72 h.

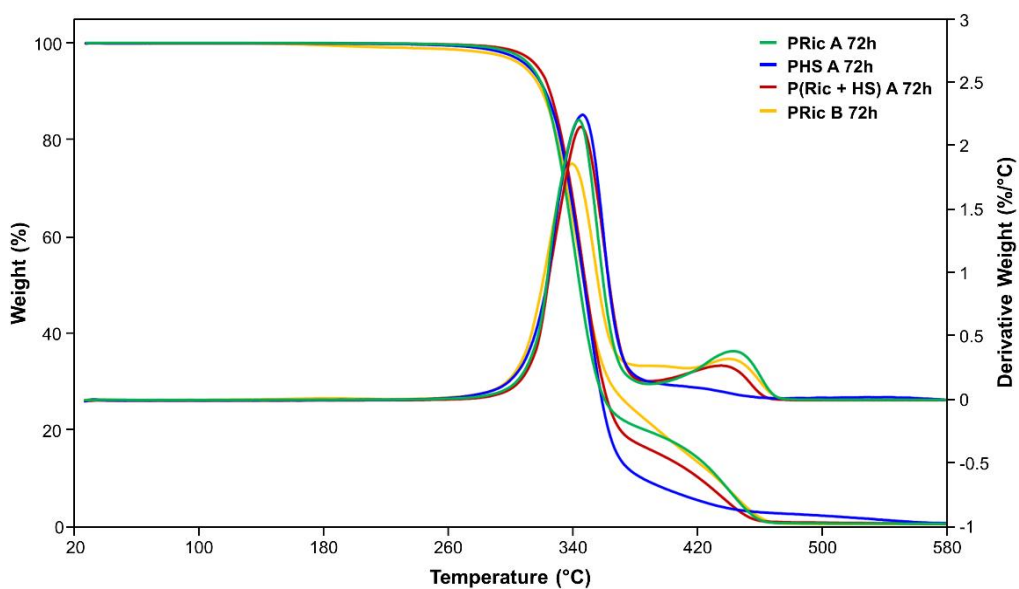

**Fig. S11.** TGA and DTG curves of high molecular weight estolides: PRic A 72 h, PHS A 72 h, P(Ric + HS) A 72 h, PRic B 72 h.

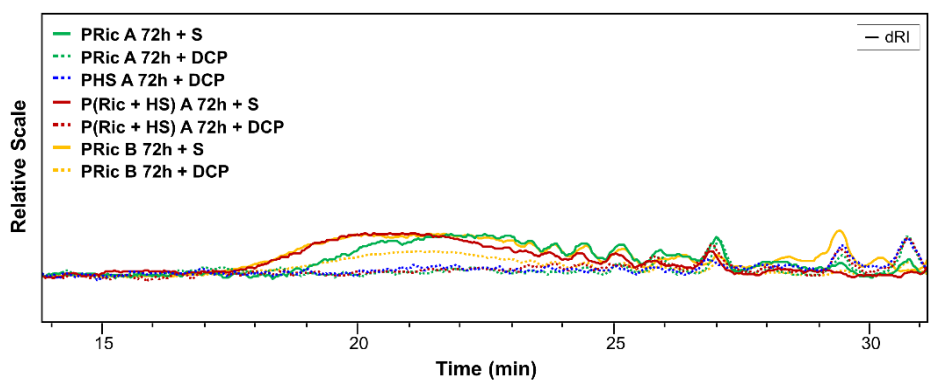

**Fig. S12.** SEC traces of the soluble fractions of cross-linked estolides: PRic A 72 h + S, PRic A 72 h + DCP, PHS A 72 h + DCP, P(Ric + HS) A 72 h + S, P(Ric + HS) A 72 h + DCP, PRic B 72 h + S, PRic B 72 h + DCP.

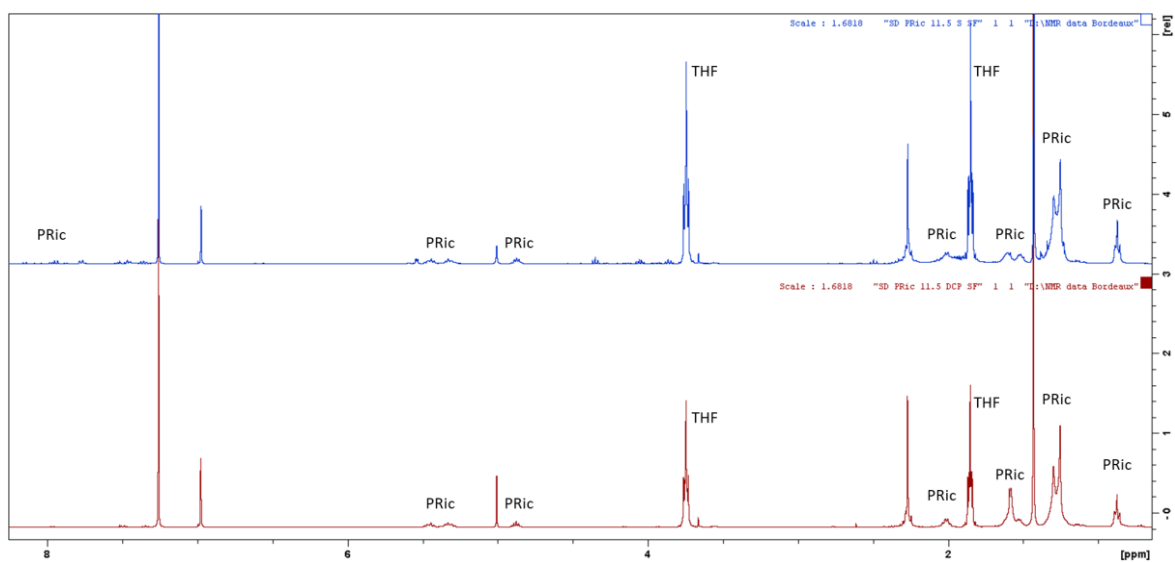

**Fig. S13.** $^1$H NMR of the soluble fraction of cross-linked estolides in CDCl$_3$: PRic A 72 h + S (blue), PRic A 72 h + DCP (red).

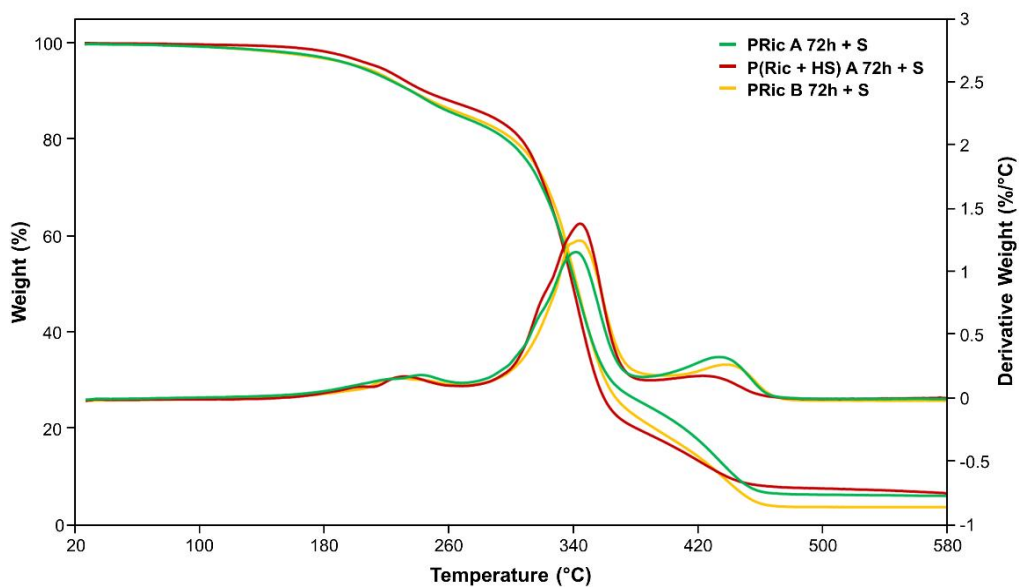

**Fig. S14.** TGA and DTG curves of high molecular weight estolides cross-linked using sulfur system: PRic A 72 h + S, P(Ric + HS) 72 h + S, PRic B 72 h + S

**Table S1.** Thermal analysis results of cross-linked estolides: PRic A 72 h + DCP, PRic A 72 h + S, PHS A 72 h + DCP, P(Ric + HS) A 72 h + DCP, P(Ric + HS) A 72 h + S, PRic B 72 h + DCP, PRic B 72 h + S.

| Cross-linked estolide | $T_{5\%}$ (°C) | Char at 500 °C (%) | $T_g$ (°C) |
|---|---|---|---|
| PRic A 72 h + DCP | 312 | 0.8 | -64 |
| PRic A 72 h + S | 203 | 6.8 | -57 |
| PHS A 72 h + DCP | 311 | 0.6 | - |
| P(Ric + HS) A 72 h + DCP | 318 | 1.5 | -56 |
| P(Ric + HS) A 72 h + S | 217 | 7.7 | -54 |
| PRic B 72 h + DCP | 260 | 1.2 | -69 |
| PRic B 72 h + S | 205 | 4.1 | -60 |